\documentclass[lnicst]{svmultln}
\usepackage{etex}
\usepackage{graphicx}
\usepackage{tabulary}
\usepackage{tabularx}
\usepackage{array}
\usepackage{makeidx} 
\usepackage{amssymb}
\usepackage{amsmath}
\usepackage{amsfonts}
\usepackage{empheq}
\usepackage{enumerate}
\usepackage{float}
\usepackage[algo2e]{algorithm2e}
\usepackage{algorithm}
\usepackage{algorithmic}
\usepackage{tikz}
\usepackage[export]{adjustbox}
\usepackage{url}
\usepackage[colorlinks=true,urlcolor=black,linkcolor=black,citecolor=black]{hyperref}
\usepackage{blindtext}
\usepackage[colorinlistoftodos,prependcaption,textsize=small]{todonotes}
\usepackage{regexpatch}
\usepackage[english]{babel}
\usepackage[normalem]{ulem}
\makeatletter
\xpatchcmd{\@todo}{\setkeys{todonotes}{#1}}{\setkeys{todonotes}{inline,#1}}{}{}
\makeatother
\usepackage{footmisc}

\usepackage{pgf-umlsd}
\usetikzlibrary{arrows, shadows, positioning}
\usepackage{makeidx}  

\usepackage{fancybox}
\usepackage{wrapfig} 
\usepackage{nameref}
\usepackage{fixltx2e}

\newcommand{\amess}[7][0]{
  \stepcounter{seqlevel}
  \path
  (#2)+(0,-\theseqlevel*\unitfactor-0.7*\unitfactor) node (mess from) {};
  \addtocounter{seqlevel}{#1}
  \path
  (#4)+(0,-\theseqlevel*\unitfactor-0.7*\unitfactor) node (mess to) {};
  \draw[->,>=angle 60] (mess from) -- (mess to) node[midway, below]
  {#3};

  \if R#5
    \node (#3 from) at (mess from) {\llap{#6~}};
    \node (#3 to) at (mess to) {\rlap{~#7}};
  \else\if L#5
         \node (#3 from) at (mess from) {\rlap{~#6}};
         \node (#3 to) at (mess to) {\llap{#7~}};
       \else
         \node (#3 from) at (mess from) {#6};
         \node (#3 to) at (mess to) {#7};
       \fi
  \fi
}

\renewcommand{\mess}[4][0]{
  \stepcounter{seqlevel}
  \path
  (#2)+(0,-\theseqlevel*\unitfactor-0.7*\unitfactor) node (mess from) {};
  \addtocounter{seqlevel}{#1}
  \path
  (#4)+(0,-\theseqlevel*\unitfactor-0.7*\unitfactor) node (mess to) {};
  \draw[->,>=angle 60] (mess from) -- (mess to) node[midway, above]
  {#3};

  \node (\detokenize{#3} from) at (mess from) {};
  \node (\detokenize{#3} to) at (mess to) {};
}

\tikzset{every picture/.append style={scale=0.9}}

\usepackage{listings}
\usepackage{color}
\definecolor{gray}{rgb}{0.4,0.4,0.4}
\definecolor{darkblue}{rgb}{0.0,0.0,0.6}
\definecolor{cyan}{rgb}{0.0,0.6,0.6}

\definecolor{dkgreen}{rgb}{0,0.6,0}
\definecolor{mauve}{rgb}{0.58,0,0.82}

\DeclareFontFamily{OT1}{pzc}{}
\DeclareFontShape{OT1}{pzc}{m}{it}{<-> s * [1.10] pzcmi7t}{}
\DeclareMathAlphabet{\mathpzc}{OT1}{pzc}{m}{it}
\newcounter{todocounter}

\begin{document}
\mainmatter  
\title{TruSDN: Bootstrapping Trust in Cloud Network Infrastructure} 
\titlerunning{Bootstrapping Trust in Cloud Network Infrastructure}
\author{Nicolae Paladi, Christian Gehrmann}
\authorrunning{Nicolae Paladi et al.} 
\institute{SICS Swedish ICT \\ Stockholm, Sweden \\ \{nicolae, chrisg\}@sics.se}
\maketitle

\begin{abstract}
Software-Defined Networking (SDN) is a novel architectural model for cloud network infrastructure, improving resource utilization, scalability and administration.
SDN deployments increasingly rely on virtual switches executing on commodity operating systems with large code bases, which are prime targets for adversaries attacking the network infrastructure.
We describe and implement $\mathsf{TruSDN}$, a framework for bootstrapping trust in SDN infrastructure using Intel Software Guard Extensions (SGX), allowing to securely deploy SDN components and protect communication between network endpoints.
We introduce \textit{ephemeral flow-specific pre-shared keys} and propose a novel defense against \textit{cuckoo attacks} on SGX enclaves.
$\mathsf{TruSDN}$ is secure under a powerful adversary model, with a minor performance overhead.
\end{abstract}

\section{Introduction}
\label{sec:Intro}

Renewed and widespread interest in virtualization -- along with proliferation of cloud computing -- has spurred a series of innovations, allowing cloud service providers to deliver on-demand compute, storage and network resources for highly dynamic workloads.
Consequently, more hardware and virtual components are added to already large networks, complicating network management.
To help address this, SDN emerged as a novel network architecture model.
Separation of the \textit{data} and \textit{control} planes is its core principle, allowing network operators to implement high-level configuration goals by interacting with a single \textit{network controller}, rather than configuring discrete network components.
The controller applies the configuration to the \textit{network edge}, i.e. to its global view of the data plane~\cite{gude:2008}.
Data and control plane separation in SDN challenges network infrastructure security best practices evolved in the decades since packet-switched digital network communication gained popularity~\cite{kreutz:2013}, \cite{paladi:2015}.

In the cloud infrastructure model, SDN allows tenants to configure complex topologies with rich network functionality, managed by a network controller.
The availability of a global view of the data plane enables advanced controller capabilities -- from pre-calculating optimized traffic routing to managing applications that replace hardware middleboxes.
However, these capabilities also make the controller a valuable attack target: once compromised, it yields the adversary complete control over the network~\cite{porras:2015}.
The global view itself is security sensitive: an adversary capable of impersonating network components may distort a controller's global view and influence network-wide routing policies~\cite{hong:2015}.

\textit{Virtual switches} are another category of security sensitive components in SDN deployments.
They execute on commodity operating systems (OS) and are often assigned the same trust level and privileges as hardware switches -- specialized network components with compact embedded software~\cite{qazi:2013} or application-specific integrated circuits.
Commodity OS are likely to contain security flaws which can be exploited to compromise virtual switches.
For example, their configuration can be modified to disobey the protocol, breach network isolation and reroute traffic to a malicious destination or compromise other network edge elements through lateral attacks.
Such risks are accentuated by the extensive control a cloud provider has over the infrastructure of its tenants.


Security and isolation of tenant infrastructure can be strengthened by confining select SDN components to trusted execution environments (TEE) and attesting their integrity before provisioning security-sensitive data. 
TEEs with strong security guarantees can be built using SGX, a set of recently introduced extensions to the x86 instruction set architecture and related hardware~\cite{anati:2013,mckeen:2013}.
Earlier work used SGX to protect computation in cloud environments, by executing modified OS instances in SGX enclaves~\cite{baumann:2014} or a data processing framework in a set of SGX enclaves~\cite{schuster:2015}.
However, while both of the above efforts \textit{highlighted} the need to secure network communication, they did not \textit{address} it.



\subsection{Contribution}
\label{subsec:contrib}
This paper makes the following contributions:
	\begin{itemize}
		\item We present $\mathsf{TruSDN}$, a framework to bootstrap trust in SDN infrastructure.
		\item We introduce flow-specific pre-shared keys for communication protection.
		\item We propose a defense against cuckoo attacks~\cite{parno:2008}, based on properties of the enhanced privacy ID (EPID) scheme~\cite{brickell:2012} used for remote enclave attestation.
		\item We describe the implementation and a performance evaluation of $\mathsf{TruSDN}$.
	\end{itemize}

\subsection{Organization}
\label{subsec:organization}
We introduce the system model in Section~\ref{sec:system_model}, describe the adversary model in Section~\ref{sec:adversary_model} and the design of $\mathsf{TruSDN}$ in Section~\ref{sec:trusdn}.
In Section~\ref{sec:security_analysis} we provide a security analysis, describe the prototype implementation and performance evaluation in Section~\ref{sec:implementation_evaluation} and review the related work in Section~\ref{sec:related}.
We discuss future work in Section~\ref{sec:limitations_of_trusdn} and conclude in Section~\ref{sec:conclusion}.

\section{System Model}
\label{sec:system_model}
In this section we describe the SDN architectural model and the SDN deployment layers.
Furthermore, we describe the use of TEEs based on Intel SGX.

\subsection{Software Defined Networking}
\begin{figure}[b!]
    \centering
    \begin{minipage}{.5\textwidth}
		\centering
		\includegraphics[width=1\textwidth]{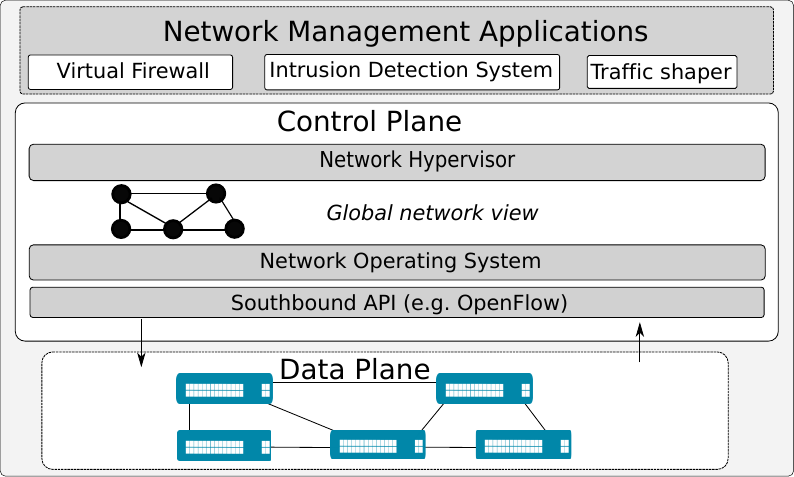}
		\caption{\small The SDN architectural model.}
		\label{fig:SDN_architecture}
	\end{minipage}%
	\begin{minipage}{.5\textwidth}
        \centering
		\includegraphics[width=0.7\textwidth]{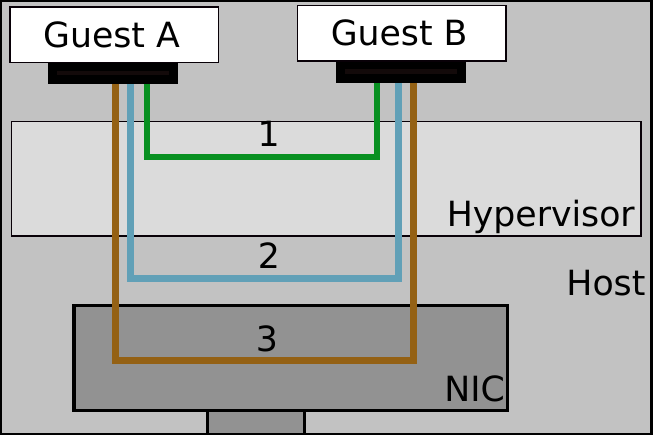}
		\caption{\small Communication paths between collocated endpoints:  
				(1) virtual switch; 
				(2) host-local, e.g. native bridging; 
				(3) virtual queues in the NIC. 
				}
		\label{fig:hairpin}
    \end{minipage}
\end{figure}

In this paper we target SDN in infrastructure cloud deployments.
The system model follows the architecture presented in \cite{casado:2014} and depicted in Figure~\ref{fig:SDN_architecture}.

The \textit{data plane} includes hardware and software switch implementations.
\textit{Software switching} is used in cloud deployments due to its scalability and configuration flexibility.
Figure~\ref{fig:hairpin} illustrates the software switching approaches for communication between two collocated endpoints.
In a typical switch implementation, its kernel-space component is optimized for forwarding performance, lacks decision logic and only forwards packets matching rules in its \textit{forwarding information base} (FIB)~\cite{nadeau:2013}.
The FIB comprises packet forwarding rules deployed to satisfy network administrator goals.
Mismatching packets are discarded or redirected to the \textit{control plane} through the \textit{southbound API}.
While the data plane uses complementary functionality of both virtual and physical switches, the role of the latter is often reduced to routing IP-tunneled traffic between hypervisors~\cite{pfaff:2015}.
In this paper we do \textit{not} address control of hardware switches and traffic routing between hosts; 
we assume that the physical network provides uniform capacity across hosts, based on e.g. equal-cost multi-path routing~\cite{hopps:2000}, such that if multiple equal-cost routes to the same destination exist, they can be discovered and used to provide load balancing among redundant paths.
Overlay networks -- e.g. VLANs or GRE~\cite{farinacci:1994} -- are used for communication between endpoints.
In this work, we focus exclusively on software switching and use the term ``switch'' to denote a virtual, software implementation.
We refer to hardware switch implementations as ``hardware switches''.

In the \textit{control plane}, high-level network operator goals are translated into discrete routing policies based on the \textit{global network view}, i.e. a graph representation of the virtual network topology.
The main component of a control plane is the \textit{network controller}, which we define as follows:
\begin{definition}
Network Controller (NC) is a logically centralized component that manages network communication in a given deployment by updating the FIB with specific forwarding rules.
The NC compiles forwarding rules based on three inputs: the dynamic global network view, the high-level configuration goals of the network operator, and the output of the network management applications.
\end{definition}
The NC is typically implemented as part of a logically centralized \textit{network OS}, which builds and maintains the global network view and may include a \textit{network hypervisor}, to multiplex network resources among distinct virtual network deployments.

\textit{Southbound API} is a set of vendor-agnostic instructions for communication between data and control planes.
It is often limited to flow-based traffic control of the data plane, with management done through a configuration database~\cite{pfaff:2015}.

Network operators use \textit{network management applications} (NMAs), e.g. firewalls, traffic shapers, etc., to configure the network using high-level commands.

\subsection{Deployment layers}
\begin{figure}[b!]
    \centering
    \begin{minipage}{.5\textwidth}
		\centering
		\includegraphics[width=0.9\textwidth]{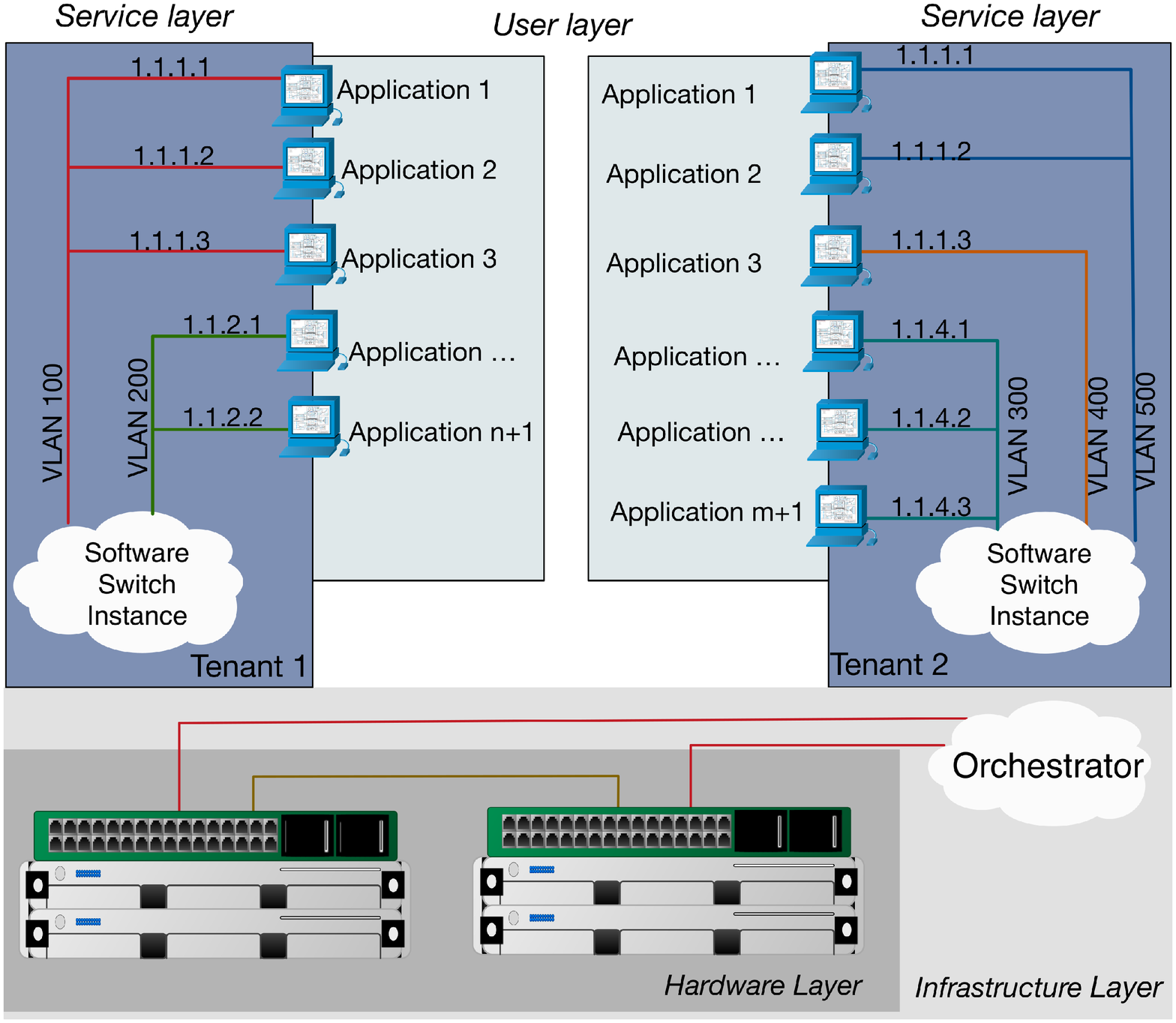}
		\caption{\small Deployment Layers.}
		\label{fig:deployment-layers}
	\end{minipage}%
	\begin{minipage}{.5\textwidth}
        \centering
\includegraphics[width=0.7\textwidth]{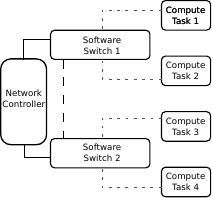}
		\caption{\small Logical communication segments:
				$\alpha$: between the NC and switches;
				$\beta$: among the switches on each host;
				$\gamma$ between host-local switches and network endpoints.
				 }
		\label{fig:logical-segments}
    \end{minipage}
\end{figure}

We next describe the deployment layers of SDN infrastructure (Figure~\ref{fig:deployment-layers}).

The \textit{hardware layer} includes infrastructure for data transfer, processing and storage and is comprised of network hardware (including hardware switches and communication channels), hardware server platforms and data storage.

The \textit{infrastructure layer} includes software components for virtualization and resource provisioning to infrastructure users, referred to as \textit{tenants}.
For network resources, this layer includes the network hypervisor, which creates \textit{network slices} by multiplexing physical network infrastructure between tenants.
Infrastructure providers expose a \textit{slice} (i.e. a quota) of network resources to the tenants.

The \textit{service layer} includes components controlled by tenants.
Network components operated by tenants are grouped into \textit{network domains}, comprising the virtual network resources and topologies that logically belong to the same organizational unit and network slice, and perform related tasks or provide a common service.
The \textit{network hypervisor} ensures that a tenant's control plane can only control switches in its own slice.
Within their slice, tenants have exhaustive creation, destruction and configuration privileges over components, such as instances of switches, the NC, NMAs and network domains.
We define three \textit{logical} communication segments (Figure~\ref{fig:logical-segments}): 
between the network controller and switches ($\alpha$ segments);
among the switches on each host ($\beta$ segments);
between host-local switches and network endpoints ($\gamma$ segments).

The \textit{user layer} includes endpoint consumers of network services, e.g. virtualization guests, containers and applications in a network domain.

\subsection{Trusted Execution Environments}
\label{subsec:tee}
The proposed solution relies on TEEs that \textit{both} provide strong isolation and allow remote code and data integrity attestation. 
Such a TEE can be created using Intel SGX enclaves (introduced in~\cite{anati:2013,mckeen:2013}) during OS runtime and relies for its security on a trusted computing base (TCB) of code and data loaded at build time, processor firmware and processor hardware.
At build time, the CPU measures the loaded code, data and memory page layout.
At initialization time, the CPU produces a final measurement, after which the enclave becomes immutable and cannot be externally modified.
The CPU maintains the measurement throughout the enclave's lifetime to later assert the integrity of the enclave contents.
Processor firmware is the root of trust (ROT) of an enclave.
It prevents access to the enclave's memory segment by either the platform OS, other enclaves, or other external agents.
Enclaves operate in a separate memory region inaccessible to non-enclave processes, called the enclave page cache (EPC).
Multiple mutually distrusting enclaves can operate on the platform.
The processor enforces separation of memory access among enclaves based on the layout in the \textit{EPC map}.
Program execution within an enclave is transparent to both the underlying OS and other enclaves.

\textit{Remote attestation} allows an enclave to provide integrity guarantees of its contents~\cite{anati:2013}.
For this, the platform produces an attestation assertion with information about the identity of the enclave and details of its state (e.g. the mode of the software environment,
associated data, and a cryptographic binding to the platform TCB making the assertion).
For \textit{intra-platform attestation} (i.e. between enclaves on the same platform), the reporting enclave (\textit{reporter}) invokes the \texttt{EREPORT} instruction to create a \texttt{REPORT} structure with the assertion and calculate a message authentication code (MAC), using a \textit{report key}, known only to the target enclave (\textit{target}) and the CPU.
The structure contains a \texttt{user data} field, where the reporter can store a hash of the auxiliary data provided.
The target recomputes the MAC with its report key to verify the authenticity of the structure, and compares the hash in the \texttt{user data} with the hash of the auxiliary data, to verify its integrity.
Enclaves then use the auxiliary data to establish a secure communication channel.
For \textit{inter-platform attestation} the remote verifier first sends a challenge to the enclave platform, where the challenge is complemented with the indentity of a \textit{quoting enclave} (QE) and forwarded to the reporter, which appends the challenge response to the \texttt{REPORT} and attests itself to the QE.
The QE verifies the structure, signs it with a platform-specific key using the \textit{enhanced privacy ID group signature scheme} (EPID)~\cite{brickell:2012} and returns it to the verifier, to check the authenticity of the signature and the report itself~\cite{anati:2013}.
The use of the EPID scheme is part of the SGX implementation and allows to maintain the privacy of the platform which hosts the enclave.

\section{Adversary Model}
\label{sec:adversary_model}
We now describe the adopted adversary model, as well as the core security assumptions on which we base our design.
The adversary model we adopt can be described by the capabilities of the adversary at the \textit{network} and \textit{platform} levels respectively
(overview in Table~\ref{tab:adv-cap}).

\subsection{Network infrastructure}
For SDN infrastructure, we adopt the adversary model introduced in~\cite{dolev:1983} and extended with SDN-specific attack vectors in~\cite{paladi:2015}.
We assume a powerful adversary ($\mathpzc{Adv}$), which controls the cloud deployment network infrastructure; 
it can intercept, record, forge, drop and replay any message on the network, and is only limited by the constraints of the employed cryptographic methods.
Particularly, the $\mathpzc{Adv}$ may forge messages that do not match any of the rules installed in the FIB.
Furthermore the $\mathpzc{Adv}$ may create own instances of switches and launch Sybil attacks~\cite{douceur:2002} and launch other types of topology poisoning attacks~\cite{hong:2015} to distort the global network view.
Finally, $\mathpzc{Adv}$ can store arbitrary quantities of intercepted communication and attempt its decryption with encryption keys intercepted or leaked at a later point.
It can analyze the traffic patterns in the network through passive probing and may disrupt or degrade network connectivity to achieve its goals.
We explicitly exclude Denial-of-Service attacks on the SDN infrastructure.

\begin{table}[b!]
 \scriptsize 
\begin{center}
\caption{Summary of the $\mathpzc{Adv}$ capabilities in relation to the adversary model.}
\label{tab:adv-cap}
  \begin{tabulary}{\textwidth}{|L| L| L|}
  	\hline
		\scriptsize \textit{Type}	 	& \textit{Network}	& \textit{Platform}  \\
    \hline
		Included 	&   \scriptsize \vtop{\hbox{\strut Intercept, record, forge, drop,}\hbox{\strut replay messages;}\hbox{\strut Analyze the traffic patterns;}\hbox{\strut Disrupt or degrade network connectivity;}\hbox{\strut Launch topology poisoning attacks}}   
			& 	\scriptsize \vtop{\hbox{\strut Control non-processor hardware;}\hbox{\strut Control software stack OS, hypervisor;}\hbox{\strut Pause execution;}\hbox{\strut Deploy arbitrary software components;}\hbox{\strut ``Cuckoo attack'': Forward function calls}\hbox{\strut to compromised SGX enclaves;}\hbox{\strut Return arbitrary values to system calls}} \\
		\hline     
		\vtop{\hbox{\strut Not included,}\hbox{\strut mitigations known}} 	&	& \vtop{\hbox{\strut Side-channels: cache-collision,}\hbox{\strut controlled channel;}\hbox{\strut Attacks on shielded execution;}}\\
		\hline
		Excplicitly excluded &  Denial-of-Service (DoS) attacks & Side-channels: power analysis; DoS attacks\\		
		\hline
  \end{tabulary}  
\end{center}
\end{table}

\subsection{Platform}
For platform security, we consider a powerful adversary, similar to~\cite{baumann:2014,schuster:2015}, that may control the entire software stack in the cloud provider's infrastructure.

On the hardware level, we assume the processor is correctly implemented and remains uncompromised; 
furthermore, we  assume a reliable and secure source of random numbers (which can be provided by the CPU).
$\mathpzc{Adv}$ has full control over the remaining hardware, including memory, I/O devices, periferials, etc.
Similarly, $\mathpzc{Adv}$ fully controls the software stack, including the platform OS and the hypervisor.
This implies that $\mathpzc{Adv}$ may pause indefinitely the execution of the code in the TEE and return arbitrary values in response to OS system calls.
However, a deployment orchestrator and NC execute under tenant control, on a fully trusted platform and software stack.
We exclude side-channel attacks.
While some side-channel attacks -- e.g. timing, cache-collision, controlled channel attacks -- can be mitigated through software modification~\cite{xu:2015},
preventing other side-channel attacks -- such as power analysis -- requires hardware modifications. 
An $\mathpzc{Adv}$ with advanced capabilities may leverage its full control over the OS to utilize the class of known attacks on shielded execution;
while we do not address such attacks, they have known countermeasures~\cite{checkoway:2013,baumann:2014}.

SGX, similar to other trusted computing solutions, is vulnerable to \textit{cuckoo attacks}~\cite{parno:2008}. 
In one attack scenario, malware on the target platform forwards the messages intended for the \textit{local} SGX enclave ($SGX^E_L$) to a remote enclave under $\mathpzc{Adv}$'s physical control (\textit{malicious} enclave, $SGX^E_M$).  
Having physical access to $SGX^E_M$, $\mathpzc{Adv}$  can apply hardware attacks to violate its security guarantees.  
As a result, $\mathpzc{Adv}$ controls all communication between the verifier and $SGX^E_L$, with access to an oracle that provides all of the answers a benign $SGX^E$ would, but without its expected security properties.

Briefly, the adversary model for platform security largely matches the remote administrator capabilities of an infrastructure cloud provider.

\section{Solution Description}
\label{sec:trusdn}
In this section we present $\mathsf{TruSDN}$, a framework for bootstrapping trust in SDN deployments.
Its goal is to allow tenants to securely deploy computing tasks and create virtualized network infrastructure deployments, given the adversary model defined in Section~\ref{sec:adversary_model}.
To satisfy this goal, the framework must satisfy the following set of requirements:
\begin{itemize}
	\item \textit{Authentication:} communication in the domain must the authenticated, and a secure enrollment mechanism for data plane components must be in place.
	\item \textit{Topology integrity:} the NC must be protected from network components that attempt to distort the global network view.
	\item \textit{Component integrity:} integrity of switches must be attested prior to enrollment and the cryptographic material required for their network access must be protected with a hardware ROT.
	\item \textit{Confidentiality protection of domain secrets:} network domain secrets -- such as VPN session keys -- should not be revealed to the $\mathpzc{Adv}$.
	\item \textit{Protected network communication:} network communication in the tenant domain must be confidentiality and integrity protected.
\end{itemize}

\subsection{TruSDN overview}
\label{sec:trusdn_overview}
\begin{figure}[b!]
\centering
\includegraphics[width=\textwidth]{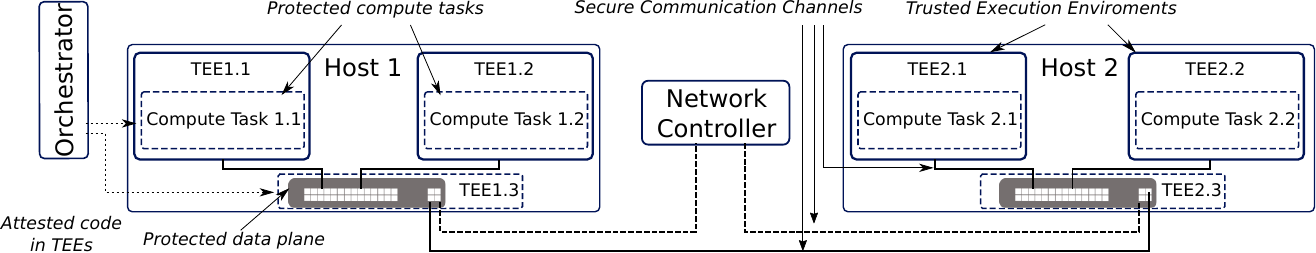}
\caption{\small Illustration of core building blocks of $\mathsf{TruSDN}$.
		 }
\label{fig:TruSDN-principles}
\end{figure}

We begin by introducing the building blocks of $\mathsf{TruSDN}$ (Figure~\ref{fig:TruSDN-principles}).
\paragraph{Trusted Execution Environments:} 
$\mathsf{TruSDN}$ uses TEEs that guarantee secure execution in the given adversary model, assuming the CPU and executed code are correctly implemented.
\paragraph{Protected Compute Tasks:}
\textit{Security sensitive} compute tasks (CT) are deployed in TEEs.
Such tasks include all operations that tenants aim to protect from the $\mathpzc{Adv}$.
However, CTs rely on the untrusted OS for I/O and support functionality.
\paragraph{Protected Data Plane:}
Switches are deployed in TEEs -- they route traffic between CTs according to forwarding rules communicated through secure channels and maintained in the FIB.
The FIB of the switches, and the key material necessary to establish the secure channels are stored in TEEs.

\paragraph{Attested code in TEEs:} 
An orchestrator under tenant control attests the TEEs during network infrastructure deployment, to ensure integrity of the deployed code and data before keys or key material are provisioned to the respective TEE. 

In a typical deployment scenario, the tenant invokes an orchestrator to deploy a switch \textit{bootstrap application} on the hosts in the tenant's domain.
The bootstrap application invokes a host-local SGX driver to build an SGX enclave containing a switch.
Next, the orchestrator attests the created enclave (as described in Section~\ref{subsec:tee}) prior to enrolling the switch with the NC.
The orchestrator uses the enclave's public key from the attestation quote to securely transfer the enclave-specific integrity and confidentiality protection session keys used to establish a protected communication channel between the NC and the TEE.
Finally, the NC communicates any remaining security-sensitive payload to the created TEE, e.g. the initial FIB.
Next, CTs are deployed in TEEs on the host and the switch forwards packets between the CTs, matching them against the rules in the FIB. 
Mismatching packets are forwarded to the NC, which may update the FIB with new rules.
For clarity, we assume the orchestrator and NC are collocated on a platform under tenant control and view both as a single component, further referred to as ``NC''.

\paragraph{Secure Communication:}
$\mathsf{TruSDN}$ protects the communication between CTs, between switches and the NC, as well as among the switches, in the above adversary model. 
Communication security is ensured using confidentiality and integrity protection keys provisioned to authenticated network components and endpoints executing in TEEs.
Furthermore, $\mathsf{TruSDN}$ leverages SDN principles to introduce a novel mechanism -- per-flow communication protection using ephemeral flow-specific pre-shared keys (PSKs).

\subsection{Cryptographic Primitives}
\label{subsec:cryptoPrimitives}
We now define the cryptographic primitives and notations used in the remainder of this paper.
We denote by $\left\{0,1\right\}^n$ the set of all binary strings of length $n$, and by $\left\{0,1\right\}^*$ the set of all finite binary strings.
In a set $U$, we refer to the $i^{th}$ element as $u_i$, and use the following notation for cryptographic operations:
\begin{itemize}
  \item Given an arbitrary message ${m \in \left\{0,1\right\}^*}$, we denote by ${c = \mathsf{Enc}\left(K, m\right)}$ a symmetric encryption of $m$ using the secret key ${K \in \left\{0,1\right\}^*}$. 
  The corresponding symmetric decryption operation is ${m = \mathsf{Dec}(K,c)=\mathsf{Dec}(K, \mathsf{Enc}(K,m))}$.
  \item We denote by $\mathsf{pk/sk}$ a public/private key pair for a public key encryption scheme. 
  We denote by  ${c = \mathsf{Enc_{pk}}\left(m\right)}$  the encryption of message $m$ with the public key $\mathsf{pk}$, and the decryption by ${m = \mathsf{Dec_{sk}}(c)=\mathsf{Dec_{sk}}(\mathsf{Enc_{pk}}(m))}$.
  \item We denote a digital signature over a message $m$ by ${\sigma = \mathsf{Sign}_{\mathsf{sk}}(m)}$ and the corresponding verification of a digital signature by ${\nu =\mathsf{Verify}_{\mathsf{pk}}(m,\sigma)}$, where ${\nu = 1}$ if the signature is valid and ${\nu = 0}$ otherwise.
  \item We denote a Message Authentication Code ($\mathsf{MAC}$) using a secret key $K$ over a message $m$ by ${\mu = \mathsf{MAC}(K,m)}$.
\end{itemize}

We next describe key sharing and communication protection mechanisms on the identified logical segments.
Table~\ref{tab:key-summary} summarizes the keys used by $\mathsf{TruSDN}$.

\begin{table}[t!]
 \scriptsize 
\begin{center}
\caption{Summary of keys used in the $\mathsf{TruSDN}$ framework.}
\label{tab:key-summary}
  \begin{tabulary}{\textwidth}{|L | C| C |C|}
  	\hline
    \textit{Key} 			& \textit{Created by}& \textit{Access}				     &	\textit{Usage} \\
    \hline
		$K_i^\alpha$ 		& NC  	  			&   NC, switch                    	 &  Enclave-specific session, segment $\alpha$\\
		$K_j^\beta$ 			& NC  		  		&   NC , switch  				     &  Domain-specific session, segment $\beta$\\
		$K'$					& NC					&   NC, switch						 &  Ephemeral session key \\
		$K''$	    			& NC					&   NC, switch						 &  Ephemeral MAC key \\
		$EK^{pk}_i$ 			& switch  			&   public                  		     &  Public key of the switch enclave\\
		$EK^{sk}_i$ 			& switch      		&   switch                  		     &  Private key of the switch enclave \\
		$CK^{pk}_i$ 			& CT  				&   public                  		     &  Public key of the compute task\\
		$CK^{sk}_i$ 			& CT			      	&   CT 			                 	 &  Private key of the compute task\\    
		$QE^{pk}$ 			& vendor   			&   public			 				 &  Public key of the quoting enclave\\
		$QE^{sk}$ 			& vendor    			&   vendor, QE 		             	 &  Private key of the quoting enclave \\
		$SK^{\gamma}_{ij}$ 	& NC 	   			&   NC, CT$_i$, CT$_j$              	 &  Ephemeral flow-specific pre-shared key \\		 		
		\hline
  \end{tabulary}  
\end{center}
\end{table}


\subsection{SDN Trust Bootstrapping and Secure Communication}
\label{sec:trust_in_sdn}

\begin{figure}[t!]
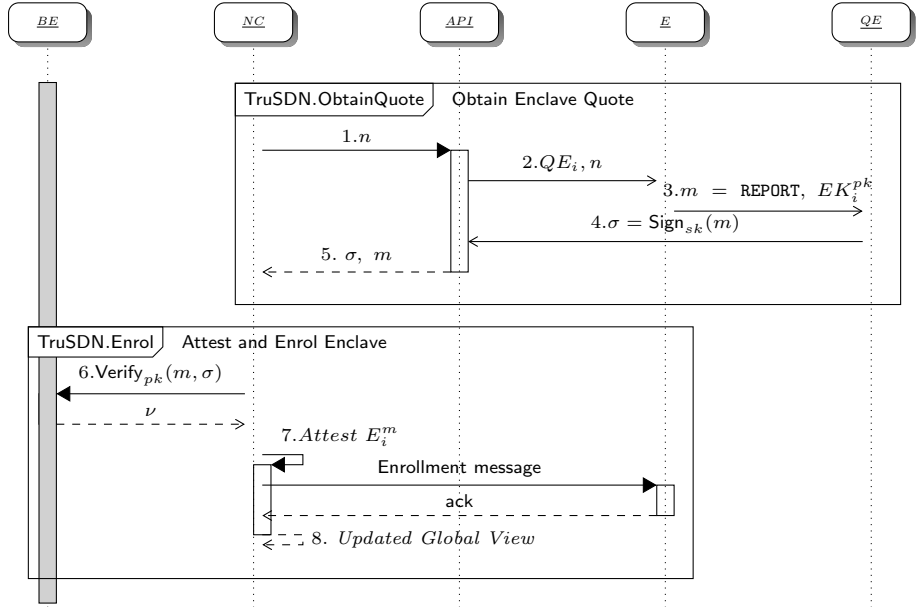

\centering
\tikzstyle{every node}=[font=\scriptsize]
	\begin{sequencediagram}
		\tikzset{inststyle/.append style={
        drop shadow={top color=gray,bottom color=white}, 
        scale=0.65,
        rounded corners=1.0ex},
        scale=0.75,
      			}
		\newthread{be}{$BE$} 
     	\newinst[2.5]{nc}{$NC$} 
     	\newinst[2.5]{api}{$API$} 
     	\newinst[2.5]{e}{$E$} 
     	\newinst[2.5]{qe}{$QE$} 
   			\begin{sdblock}{$\mathsf{TruSDN.ObtainQuote}$}{\scriptsize\textsf{Obtain Enclave Quote}}                                    
				\begin{call}{nc}{\scriptsize\textsf{$1. n$ }}{api}{$5.\ \sigma,\ m$}   
					\mess{api}{$2. QE_i, n $}{e}
					\mess{e}{$3. m\ =\ \texttt{REPORT},\ EK^{pk}_i$}{qe}
					\mess{qe}{$4. {\sigma = \mathsf{Sign}_{sk}(m)}$}{api}
   				\end{call}
      		\end{sdblock}  
   			\begin{sdblock}{$\mathsf{TruSDN.Enrol}$}{\scriptsize\textsf{Attest and Enrol Enclave}}                                    
				\begin{call}{nc}{\scriptsize\textsf{$6. \mathsf{Verify}_{pk}(m,\sigma)$}}{be}{$\nu$}   
   				\end{call}
   				\begin{callself}{nc}{$7. Attest\ E_i^m$}{$8.\ \textit{Updated\ Global\ View}$}
				\begin{call}{nc}{\scriptsize$\mathsf{Enrollment\ message}$}{e}{$\mathsf{ack}$}   
   				\end{call} 
				\end{callself}
      		\end{sdblock}   			
	\end{sequencediagram}
\caption{$\mathsf{TruSDN}$ enclave attestation and enrollment:
		 (1.) Random nonce \textit{n} is 
		 (2.) supplemented with the host QE identity;
		 (3.) Quote \textit{m} produced by the enclave is 
		 (4.) signed by the QE.
		 (6.) The verifier checks the signature of the QE, (7.) attests the integrity of the enclave and (8.) only enrolls the enclave upon success.
		 $BE$: back-end.
		 }
\label{fig:e_provision}
\end{figure}

The first step in deploying a $\mathsf{TruSDN}$ infrastructure is to launch a set of trusted switches for connectivity and topology building.
The NC requests the creation of switch enclaves to deploy switches in TEEs on hosts in its domain.
Switches are deployed based on parameters provided by the NC in plaintext (application code and configuration).
Next, the NC \textit{attests} the integrity of switch enclaves and only \textit{enrolls} the successfully attested ones 
(Figure~\ref{fig:e_provision}). 
A TEE $E_i$ is attested following the protocol introduced in~\cite{anati:2013}.
With $\mathsf{TruSDN}$ however, the reporter generates an enclave-specific public-private keypair and submits its public key $EK^{pk}_i$ along with the attestation data; 
a hash of the public key is stored in the \texttt{user data} field.
The switch enclave is only enrolled to the global network view if its reported state matches the one expected by NC.

Having attested enclave $E_i$, NC communicates an $\mathsf{Enrollment\ message}$ (Table~\ref{tab:enrol-msg}) with the 
enclave-specific pre-shared key $K_i^{\alpha}$ and 
domain-specific pre-shared key $K_j^{\beta}$, encrypted with an ephemeral key $K'_i$. 
Switches within a domain use $K_j^{\beta}$ to protect communication on $\beta$ segments. 
The $NC$ appends a MAC of the message calculated with $K''_i$ and encrypts the keys $K'_i,\ K''_i$ with $EK^{pk}_i$.

\begin{table}[b!]
 \scriptsize 
\begin{center}
\caption{$\mathsf{Enrollment\ message}$ sent by the NC upon switch enrollment.}
\label{tab:enrol-msg}
  \begin{tabulary}{\textwidth}{|L | L | C|}
  	\hline
      $m\ =\ \mathsf{Enc}(K'_i, (K_i^\alpha, K_j^\beta)) $ & ${\mu = \mathsf{MAC}(K''_i,m)}$ & $\mathsf{Enc}(EK^{pk}_i, (K'_i, K''_i)$  \\
	\hline
  \end{tabulary}  
\end{center}
\end{table}

Once switches are deployed and enrolled, tenants may configure the network topology using the NC to update the switch FIBs.
Communication on $\alpha$ segments -- e.g. FIB updates or unmatched packets forwarded to the NC --
is protected using the session key~\textit{$K_i^\alpha$} (e.g. using TLS~\cite{eronen:2005}), which never leaves the TEE.

Similarly, a secure channel is established among the switches within the same domain, using the pre-shared key~\textit{$K_j^\beta$}, to protect communication between switches on different hosts (e.g. TEEs 1.2 and 2.3 in Figure~\ref{fig:TruSDN-principles}).
$K_j^\beta$ never leaves the TEEs, has a limited validity time and is periodically redeployed by the NC.
On $\beta$ segments, traffic may traverse multiple hardware switches, forwarded to the host over tunnels deployed on top of a standard routing protocol (e.g.~\cite{hopps:2000}).

Next, the tenant may deploy CTs in TEEs and attest their integrity using the very same scheme and principles as for the switch deployment described above.
The CTs and the network controller use the $\mathsf{Enrollment\ message}$ to establish a secure communication channel (e.g. TLS).


\begin{figure}[b!]
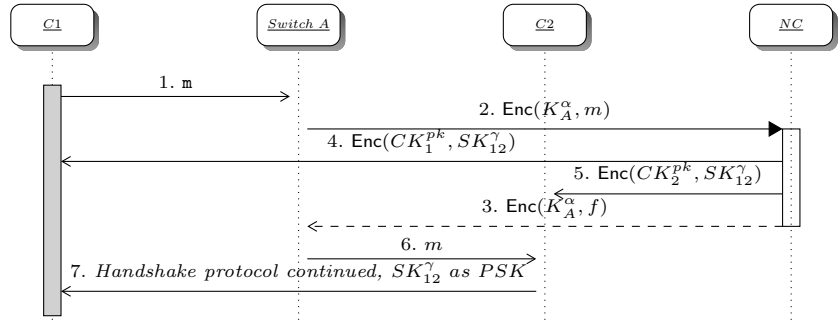

\tikzstyle{every node}=[font=\scriptsize]
	\begin{sequencediagram}
		\tikzset{inststyle/.append style={
        drop shadow={top color=gray,bottom color=white}, 
        scale=0.69,
        rounded corners=1.0ex},
        scale=0.8,
      			}
		\newthread{cc1}{$C1$}    
     	\newinst[3]{vs}{$Switch\ A$} 	 
     	\newinst[3]{cc2}{$C2$} 
     	\newinst[3]{nc}{$NC$} 	 
   					\mess{cc1}{$1.\ \texttt{m}$}{vs}
				\begin{call}{vs}{\scriptsize\textsf{$2.\ \mathsf{Enc}(K_{A}^{\alpha}, m)$}}{nc}{$3.\ \mathsf{Enc}(K_{A}^{\alpha}, f)$}   
				\mess{nc}{$4.\ \mathsf{Enc}(CK_1^{pk}, SK_{12}^{\gamma})$}{cc1}				
				\mess{nc}{$5.\ \mathsf{Enc}(CK_2^{pk}, SK_{12}^{\gamma})$}{cc2}								
				\end{call}
				\mess{vs}{\scriptsize\textsf{$6.\ m$}}{cc2}   
   				\mess{cc2}{$7.\ \textit{Handshake protocol continued, }SK_{12}^{\gamma}\ as\ PSK$}{cc1}				
    \end{sequencediagram}
\caption{Intra-host communication with $\mathsf{TruSDN}$.}
\label{fig:trusdn-intra}
\end{figure}

Once the NC has deployed and attested the TEEs with switches and CTs, intra-host communication (i.e. between two CT enclaves on the same host) is straightforward
(Figure~\ref{fig:trusdn-intra}):
when a packet $m$ sent from C1 (e.g. a TLS \texttt{ClientHello} message) reaches the local host \textit{switch A}, it attempts to match \textit{m} against a FIB entry;
if no suitable flow rule $f$ is present, the switch forwards $\mathsf{Enc}(K_A^{\alpha}, m)$ to NC, which processes the packet, generates and deploys on the CTs C1, C2 a flow-specific pre-shared key $SK_{12}^{\gamma}$ and finally updates the switch FIB with $f$, after which steps 2 and 3 are ignored;
once the FIB is updated, the switch forwards $m$ to C2, which continues the message exchange and uses $SK_{12}^{\gamma}$ to protect the communication with C1, using e.g. TLS~ with a PSK ciphersuite\cite{eronen:2005}.

Communication between CTs C1 and C3 deployed on distinct hosts is similar, with the only notable difference that the NC updates the FIB of the local switches on both hosts where C1, C3 are deployed.

In the above scenarios $\mathsf{TruSDN}$ leverages two aspects of the SDN model -- 
\textbf{(1)} the deployment has a central authority (the NC) and \textbf{(2)} the first packet of a flow is forwarded to the central authority --
to deliver on demand ephemeral PSKs to communication endpoints.
This allows to relax the need for high-quality entropy being available to CTs (a known issue in virtualized environments~\cite{ristenpart:2010}).
Furthermore, this approach ensures communication security without compromising packet visibility -- 
having control over the keys used to protect communication between the CTs allows the NC to maintain fine-grained insight into the traffic.

\subsection{Preventing Cuckoo Attacks}
\label{subsec:anti-cuckoo}
To prevent cuckoo attacks~\cite{parno:2008}, we propose a solution that leverages cryptographic properties of the EPID scheme used by the QE~\cite{brickell:2012} and the \textit{SIGn and Message Authentication} (SIGMA) protocol~\cite{walker:2011}, which are both part of the Intel SGX implementation.
The EPID scheme supports two signature modes:
\textit{fully anonymous mode} -- the verifier cannot associate a given signature with a particular member of the group; 
\textit{pseudo-anonymous mode} -- the verifier can determine whether it has verified the platform previously.
The unlinkability property distinguished in the two modes depends on the chosen \textit{base}.
A signature includes a pseudonym $\mathsf{B^f}$, where $\mathsf{B}$ is the base chosen for a signature and revealed during the signature; 
$\mathsf{f}$ is unique per member and private.
For a \textit{random base} $\mathsf{R}$, the pseudonym is $\mathsf{R^f}$ -- in this case the signatures are unlinkable.
For a \textit{name base}, the pseudonym is $\mathsf{N^f}$, where $\mathsf{N}$ is the name of the verifier --
in this case  the signatures remain unlinkable for \textit{different} verifiers, while signatures with a common $\mathsf{N}$ can be linked.
For privacy reasons, the EPID scheme currently implemented in Intel SGX accepts \textit{name base} pseudonyms only from verifiers authorized by the EPID authority~\cite{ruan:2014}, which is done by provisioning qualified verifiers with an X.509 certificate -- e.g. an intermediate certification authority (CA) certificate -- signed by the EPID authority acting as root CA.

We propose the following algorithm to prevent cuckoo attacks.
At deployment time, the EPID authority issues, to an authorized verifier $V_P$, an intermediate CA verifier certificate for the platforms in the cloud provider's data center.
Next, $V_P$ attests its platforms following the SIGMA protocol and publishes a list of resulting platform EPID signatures and the signature name base, $\mathsf{B^N_P}$.
To guard against cuckoo attacks, tenants first request $V_P$ to issue an X.509 certificate and enable them to become \textit{authorized verifiers}.
Next, tenants choose the same pseudonym base $\mathsf{B^N_P}$ (and a private $\mathsf{f}$), follow the SIGMA protocol, and verify that the resulting signature is linkable to a signature in the published list.
The cloud provider has multiple tools to protect platform privacy and prevent untrusted tenants from fingerprinting the platform infrastructure, e.g. limiting the validity of issued certificates, changing the name base, etc.
Considering that the EPID scheme is currently not implemented in the SGX emulation software we used for prototyping, we intend to describe the implementation of the above algorithm in a follow-up report.

\section{Security Analysis}
\label{sec:security_analysis}

In this section we analyze the security properties of the proposed framework in the adversary model described in Section~\ref{sec:adversary_model}.
On the network level, many of the $\mathpzc{Adv}$ capabilities are thwarted by first authenticating the switches deployed on the data plane, as well as the network edge (i.e. the compute tasks that generate or receive the network traffic), in combination with confidentiality and integrity protection of the traffic on the three identified segments.
Authenticating the network components prevents topology poisoning attacks (a countermeasure mentioned in~\cite{hong:2015}), while confidentiality and integrity protection of all of the network traffic in the deployment prevents the  $\mathpzc{Adv}$ from either learning the contents of the exchanged packets or successfully forging packets.
The $\mathpzc{Adv}$ may in this case still intercept and record messages.
However, collecting encrypted traffic does not yield the $\mathpzc{Adv}$ any more information about the contents of the exchanged packets.
Similarly, the $\mathpzc{Adv}$  does not gain an advantage by simply dropping or replaying messages, since these actions would at most simply reduce the channel capacity (as would the ability of the $\mathpzc{Adv}$ to disrupt network connectivity).
Finally, the proposed framework does not prevent the $\mathpzc{Adv}$ analyzing the traffic patterns and does not prevent it from fingerprinting the components of the deployment, making it vulnerable to rule scanning and denial of service attacks.
While the goals of $\mathsf{TruSDN}$ did not include this, such traffic analysis could be prevented using anti-fingerprinting techniques, as proposed in~\cite{bifulko:2015}.

On the platform level, the security of the proposed framework relies to a large extent on the security properties of Intel SGX enclaves.
This allows to protect the execution of switches and network edge components deployed in TEEs from the capabilities of an $\mathpzc{Adv}$ controlling non-processor hardware, the software stack of the OS and the hypervisor.
Similarly, pausing execution of switches executing in TEEs, while possible, would have no further effect than degrading network connectivity, already discussed above.
While the $\mathpzc{Adv}$ may attempt to deploy own arbitrary components on the data plane or the network edge in order to launch Sybill attacks, the integrity of such components would not be successfully attested, unless they are identical to legitimate components, which are assumed to be executing correctly -- rendering Sybill behavior impossible.
The $\mathpzc{Adv}$ is prevented from launching cuckoo attacks by enabling tenants to verify the platforms, as described in Section~\ref{subsec:anti-cuckoo}.
As presented in Table~\ref{tab:adv-cap}, several relevant classes of attacks are not addressed by $\mathsf{TruSDN}$, but have known mitigations, namely cache-collision, controlled channel and attacks on shielded execution (addressed in~\cite{xu:2015, schuster:2015}).
The capability of the $\mathpzc{Adv}$ to return arbitrary values to system calls, while not addressed in this work, can be mitigated by a validation component as described in~\cite{baumann:2014}.

\section{Implementation and Evaluation}
\label{sec:implementation_evaluation}
We now describe the implementation and evaluation of $\mathsf{TruSDN}$.

\subsection{$\mathsf{TruSDN}$ Implementation}
\label{subsec:trusdn-impl}
\begin{figure}[t!]
\centering
\includegraphics[width=0.7\textwidth]{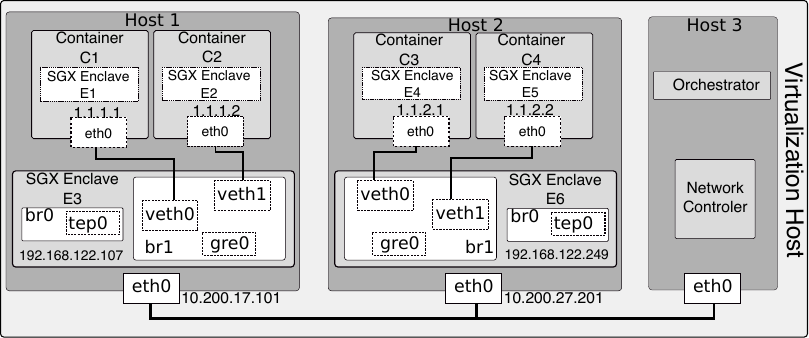}
\caption{\small Prototype deployment of $\mathsf{TruSDN}$}
\label{fig:TruSDN-arch}
\end{figure}

The $\mathsf{TruSDN}$ prototype deployment follows the design presented in Section~\ref{sec:trusdn} and is illustrated in Figure~\ref{fig:TruSDN-arch}.
\textit{Host~1} and \textit{Host~2} are instances of Ubuntu OS 15.04.
In each instance, we deployed Linux Containers\footnote{Linux Containers Project Website: \url{https://linuxcontainers.org/}}, similarly based on Ubuntu OS 15.04.
Containers create an environment with own process and network space, implemented using \textit{namespaces}, with a distinct user ID, network stack, mount points, file systems, processes, inter-process communication, and hostname.
We chose containers to facilitate prototype implementation, using their lightweight process isolation.
Containers are part of the untrusted OS and this implementation choice is orthogonal to the security of $\mathsf{TruSDN}$.
Compute tasks are deployed in TEEs created using  SGX enclaves (Figure~\ref{fig:TruSDN-arch}): \textit{enclaves} E1, E2, E4, E5 are placed respectively within containers C1, C2, C3, C4.
The switches are deployed in TEEs created using SGX enclaves (enclaves E3, E6 in Figure~\ref{fig:TruSDN-arch}).

Considering that platforms with hardware and software support for SGX were not publicly available at the time of writing, we used \textit{OpenSGX}~\cite{prerit:2016} to emulate the TEEs.
It is a software SGX emulator and  a platform for SGX development, implemented using binary translation of QEMU and emulating Intel SGX hardware components at instruction level. 
It includes emulated hardware and OS components, enclave program loader, the OpenSGX user libraries, debugging and performance monitoring support. 
The emulator allows to implement, debug, and evaluate SGX applications, but does not support binary compatibility with Intel SGX.
Furthermore, OpenSGX  does not implement all instructions, e.g. debugging instructions.
While OpenSGX does not provide security gurantees, it allows us to obtain performance estimates for the proposed approach.
We used \textit{mbedTLS}\footnote{mbed TLS project website \url{https://tls.mbed.org/}} v1.3.11 (distributed with the emulator) for attestation of the SGX enclaves. 
We used OpenSSL v1.0.2d (distributed with the emulator) to set up protected communication channels between the CT enclaves and the local switches, and among switches within the same domain.

An SDN network controller is deployed in a third instance (\textit{Host~3}).
We used the \textit{Ryu}\footnote{Ryu SDN framework:~\url{https://osrg.github.io/ryu/}} SDN framework, due to its flexibility and versatile APIs.

\subsection{$\mathsf{TruSDN}$ Evaluation}
\label{subsec:trusdn_evaluation}
We now analyze the performance impact, present evaluation results and discuss aspects that cannot be measured with the current prototype.

\subsubsection{Sources of Performance Impact}
$\mathsf{TruSDN}$ introduces several potential sources of performance impact (Table~\ref{tab:ovh-sources}).
We distinguish between \textit{transient} performance overhead, which occurs occasionally (e.g. TLS key negotiation) and \textit{continuous} performance overhead, present throughout the infrastructure operation.
We \textit{do not} consider the one-time cost of infrastructure deployment, e.g. provisioning the software, attesting TEEs and enrolling the components.
\begin{table}[t]
 \scriptsize 
\begin{center}
\caption{Sources and types of performance overhead in $\mathsf{TruSDN}$}
\label{tab:ovh-sources}
  \begin{tabulary}{\textwidth}{|L | C| C |}
  	\hline
    \textit{Source} 					   			   & \textit{Type} & \textit{Clarification} \\
    \hline
		TLS negotiation all segments & transient  & Negotiate session keys for TLS \\
		PSK distribution				& transient  & Distribute PSK for $\gamma$ segments\\		
		TLS protection all segments  & continuous & Overhead induced by TLS  \\
		Compute task execution in TEEs  			   & continuous 	& Overhead induced by TEE   \\
		Switch execution in TEEs        			   & continuous & Overhead induced by TEE 	\\
		\hline
  \end{tabulary}  
\end{center}
\end{table}

\subsubsection{Measured Performance Impact}
To evaluate the performance impact, we measured the footprint of establishing TLS sessions on $\alpha$ and $\gamma$ segments.
We used \textit{iperf}, \textit{openssl s\_time} and an own Ryu application (Table~\ref{tab:ovh-stats}).

\paragraph{TLS overhead on the $\alpha$ segment:}
We measured the round-trip latency of packets sent in plaintext and with TLS, over 1000 tests, each request sending messages of 100 bytes with the 80 bit OpenFlow header.
Furthermore, we measured the data transfer rates for plaintext and TLS communication.
Use of TLS increased total transfer time by 14.2\% and reduced the transfer rate by 15.98\%.

\begin{table}[b]
 \scriptsize
\begin{center}
\caption{Summary of performance evaluation of $\mathsf{TruSDN}$}
\label{tab:ovh-stats}
  \begin{tabulary}{\textwidth}{|L | C| C | C | C | C |}
  	\hline
    \textit{Data} 									    & \textit{Minimum} & \textit{Maximum} & \textit{Mean} & \textit{Median} & \textit{Stddev} \\
    \hline
	Total transfer time, ms           					&0.4			   	   &1.1			     &0.66	    		&0.7				   & 0.07 \\
    \hline
	Total transfer time w. $\mathsf{TruSDN}$, ms			&0.5				   &7.1				 &0.8			&0.8					& 0.22  \\	
    \hline
    $\mathsf{TruSDN}$ overhead, \textbf{total transfer time}   &		 		   &					&\textbf{21.2\%}		&\textbf{14.2\%}			    &		\\
    \hline
    \hline
	Transfer rate, bytes per second	 								&1225			    &2095			  &1595			&1583				& 98.07  \\
    \hline
	Transfer rate w. $\mathsf{TruSDN}$, bytes per second				&919					&1589			 &1338			&1330				& 64.86  \\
    \hline
    $\mathsf{TruSDN}$ overhead, \textbf{transfer rate}				&		 		   &					& \textbf{16.11\%}		&\textbf{15.98\%}		    &		\\    
    \hline
    	\hline
	First packet latency $\gamma$ 							 &1.53	   &	6.50				&3.48			&3.38				& 0.42  \\
    \hline
	First packet latency $\gamma$ w. $\mathsf{TruSDN}$	 &3.35	   &10.7				&5.37			&5.14				& 0.93  \\
    \hline
    $\mathsf{TruSDN}$ overhead, \textbf{first packet latency}		&		 		   &					& \textbf{54.31\%}		&\textbf{52.07\%}			    &		\\        
    \hline    
	\hline
	TLS handshake, ms 						  		  &36.53	   &	77.72				&67.97			&67.48				& 7.42  \\
    \hline
	TLS handshake w. $\mathsf{TruSDN}$, ms	 	  	  &52.35	   &76.44				&67.15			&66.53				& 3.93  \\	
    \hline
    $\mathsf{TruSDN}$ overhead, \textbf{TLS handshake} &		 &						& \textbf{-2.21\%}&\textbf{-2.41\%}	&		\\        
    \hline 
   	\hline    
	Key generation NC, ms 						  	  &0.11				&0.51			&0.178			&0.16				& 0.04  \\
    \hline
	Key distribution $\gamma$, ms 				  	  &0.37				&1.06			&0.54			&0.53				&0.08  \\
	\hline
	Key total 		$\gamma$, ms 				  	  &0.50			&1.30			&0.71	     	&0.7					&0.11  \\	
	\hline
  \end{tabulary}  
\end{center}
\end{table}

\begin{figure}[t]
\centering
\begin{minipage}{0.49\textwidth}
\includegraphics[width=\textwidth]{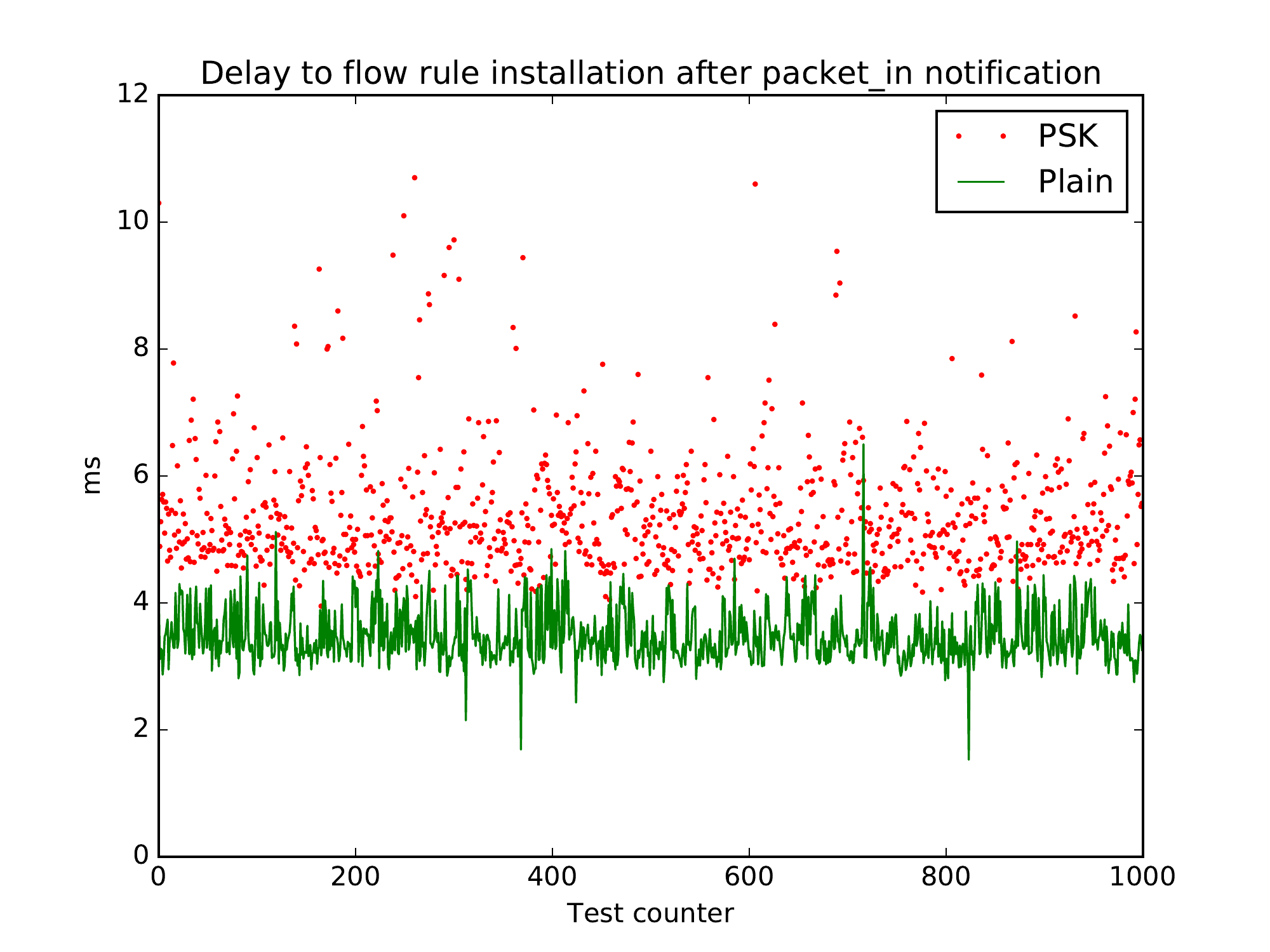}
\caption{Coarse-grained view}
\label{fig:beta-pingtest.pdf}
\end{minipage}
\begin{minipage}{0.49\textwidth}
\includegraphics[width=\textwidth]{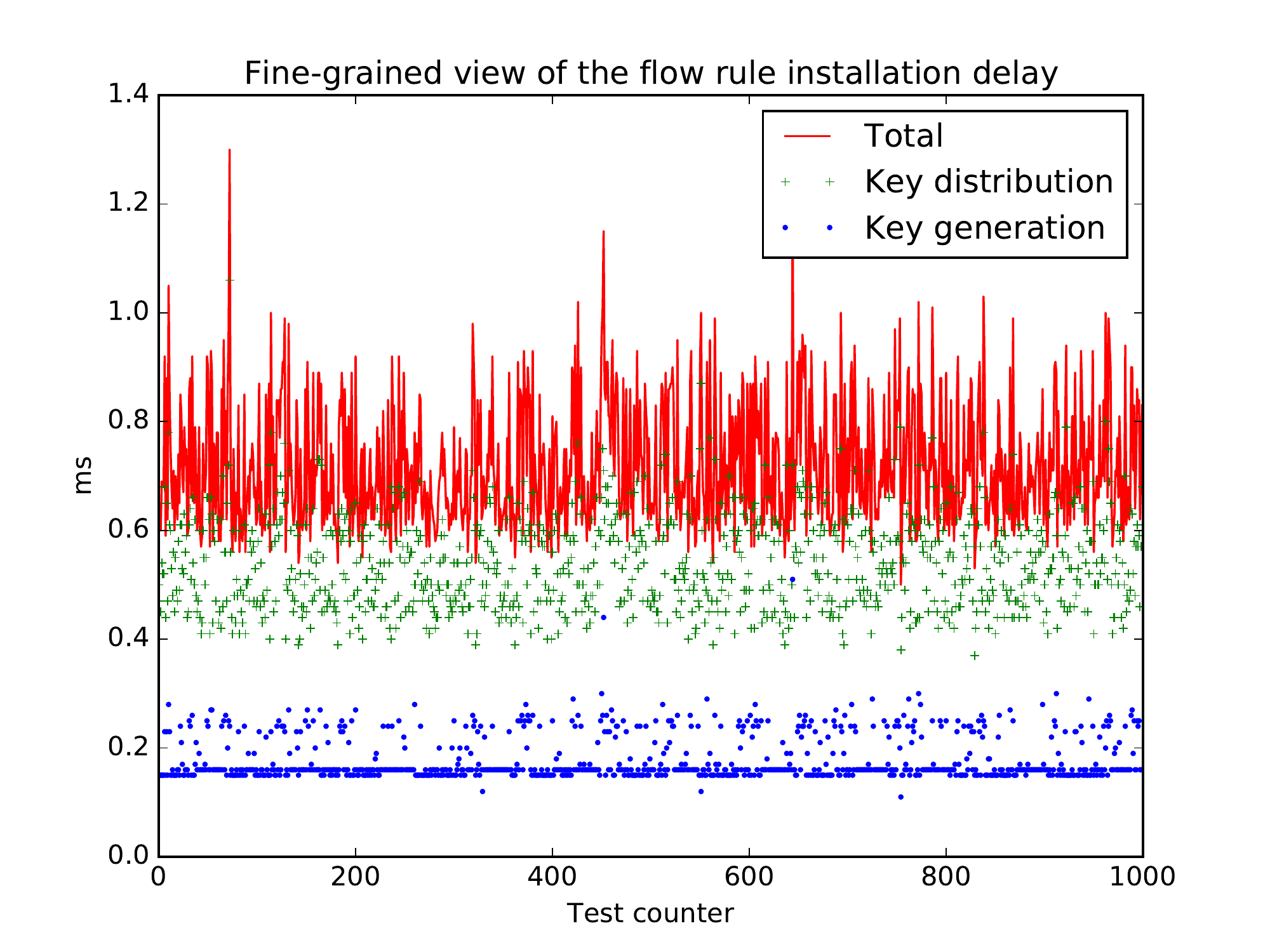}
\caption{Fine-grained view}
\label{fig:beta-keydist.pdf}
\end{minipage}
\end{figure}

\paragraph{Delay on $\gamma$ segment}
As mentioned above, the first packet of the flow is intercepted by the switch and forwarded to the NC in a \texttt{packet\_in} message~\cite{pfaff:2012}. 
At this point the NC processes the flow and installs a flow rule on the switch.
$\mathsf{TruSDN}$ \textit{extends} this procedure by generating and distributing to the communicating CTs a pre-shared key, to be used for communication protection.
Since this must be done prior to both forwarding the message to the destination CT and installing the flow rule, generating and distributing the PSK would normally delay the installation of the flow rule and increase the latency of the \textit{first} packet (all subsequent packets are forwarded according to the flow rule).
To measure the introduced delay, we have sequentially established 1000 TLS sessions between compute tasks C1 and C2 (according to Figure~\ref{fig:TruSDN-arch}).
After each TLS session, we flushed the installed flow rules (with \texttt{ovs-ofctl del-flows br0}), which resulted in a \texttt{packet\_in} message upon each new session.
The latency of the first packet is shown in Figure~\ref{fig:beta-pingtest.pdf}, and compared against the latency of a first packet without the $\mathsf{TruSDN}$ extension.

The induced delay is primarily caused by two operations performed by the NC: \textit{generating} a 256-bit PSK and distributing it to the CTs.
Figure~\ref{fig:beta-keydist.pdf} displays a fine-grained picture of the induced delay.
Key generation lasted on average 0.178 ms, while key distribution on average 0.54 ms (Table~\ref{tab:ovh-stats}).
We remind that the test environment is fully virtualized and posit that overhead of key generation can be reduced in a production environment, either by using pre-generated keys or with specialized hardware (e.g. crypto processors). 
In our tests, the duration of establishing a TLS session with ephemeral flow-specific pre-shared keys using the PSK-AES256-CBC-SHA cipher suite was 2.41\% \textit{less} compared to the use of e.g. ECDH-RSA-AES128-SHA256.
Thus, $\mathsf{TruSDN}$ enables flexible use of pre-shared keys, which in turn reduces the duration of the TLS handshake, by avoiding expensive public key cryptographic operations~\cite{kuo:2006}.
Moreover, it reduces the CPU utilization for key derivation in CTs, at the cost of a minimal flow rule installation delay.
The above approach may be applicable to other protocols.
For example, none of the differences between the datagram TLS (DTLS) and TLS protocols specified in~\cite{rfcdtls:2012} indicate that the above approach is incompatible with DTLS.
We leave further investigation for future work.

\subsubsection{Unmeasured Performance Overhead}

Implementing TEEs with OpenSGX limits the level of detail when it comes to performance evaluation, since: 
\textbf{(a)} the OpenSGX emulator \textit{is not} binary compatible with Intel SGX~\cite{prerit:2016};
\textbf{(b)} in its current version\footnote{Commit e0713c7 on \url{https://github.com/sslab-gatech/opensgx}} and unlike Intel's description of SGX~\cite{anati:2013}, OpenSGX has yet to implement support multithreaded applications\footnote{Issue \#34 on \url{https://github.com/sslab-gatech/opensgx/issues/34}}.
Thus, a fully accurate measurement on TruSDN performance cannot be done until Intel SGX hardware and software is made available. 
However, we believe our experiments yield a fair picture of the expected performance impact.

\section{Related work}
\label{sec:related}

\paragraph{Adversary models:}
Kreutz et al. presented a list of attack vectors in SDN~\cite{kreutz:2013} (forged traffic flows, vulnerabilities in switches and NCs, lack of trust establishment mechanisms, etc.). 
However, only part of the described attack vectors are exclusively relevant to SDN networks and no specific solutions are proposed.
Work in~\cite{paladi:2015} introduced an adversary model, attack vectors, and security requirements towards multi-tenant SDN infrastructure, highlighting the need to limit the effect of NC vulnerabilities, protect internal SDN communication, verify integrity of SDN components prior to enrollment, and enforce policy and quota isolation.
$\mathsf{TruSDN}$ addresses several of the attack vectors described in~\cite{kreutz:2013,paladi:2015}.

\paragraph{Secure SDN controllers:}
The ``NOX'' network OS~\cite{gude:2008} presents NMAs with a centralized programming model, allowing to operate with higher-level abstractions and apply graph processing algorithms to compute paths.
It consists of several controller processes which use the global view for network management decisions and update switch FIBs over the OpenFlow API~\cite{mckeown:2008}.
FortNOX~\cite{porras:2012} extends NOX with role-based authorization (RBA) and enforcement of security constraints.
It translates high-level threats into flow rules to handle suspicious traffic as well as detects rule conflicts, resolves them depending on the authorization of the rule requestor and enforces least privilege authorization.
Neither NOX nor FortNOX address malicious network components and Sybill attacks, addressed by $\mathsf{TruSDN}$.
``Rosemary'' NOS~\cite{shin:2014} uses NMA sandboxing to improve network resilience, by launching each NMA in a separate process context with access to the required libraries, along with a resource monitor to supervise NMA compliance.
It does not address data plane security; $\mathsf{TruSDN}$ complements it and creates a foundation for trusted deployment of a secure NOS.
TopoGuard~\cite{hong:2015} detects network topology poisoning and mitigates this through port property management, network edge probing and verification of topology updates.
$\mathsf{TruSDN}$ complements this by verifying the \textit{integrity} of switches prior to enrollment into the topology.

\paragraph{Software Guard Extensions:}
SGX was introduced in~\cite{mckeen:2013} with a description of the software model, extensions to the x86 ISA and hardware modifications for isolated execution;
work in~\cite{anati:2013} described CPU based attestation.
SGX-based solutions in a cloud setting are first described in~\cite{baumann:2014,schuster:2015}.
``Haven''\cite{baumann:2014} is a modified version of Windows 8 OS ported to an SGX enclave, evaluated with Apache Web Server and SQL Server using synthetic data sets.
It includes a mechanism to protect the enclave from a malicious kernel and a semantically secure data store protecting data and file metadata confidentiality against malicious hosts.
$\mathsf{TruSDN}$ protects network communication for a similar adversary model.
While we deploy compute tasks in SGX enclave-based TEEs, the work in~\cite{baumann:2014} is largely complementary, and similar ``Haven''-like OSs could be used.

``VC3''~\cite{schuster:2015} is a Map-Reduce deployment using SGX enclaves.
\textit{Map} and \textit{reduce} functions are compiled into private (encrypted) code and public code implementing key exchange and job execution protocols.
Code is initialized in enclaves and attested by the users.
Public code performs the key exchange, decrypts the private code and runs the job execution protocol.
To defend against cuckoo attacks, \textit{cloud quoting enclaves} are created on \textit{each} platform in the cloud provider data centers, to ``countersign'' quotes produced by the QE.
The approach is largely complementary to protecting communication between CTs with $\mathsf{TruSDN}$.
However, the proposed defense against cuckoo attacks increases the complexity of the attestation protocol and does not prevent $\mathpzc{Adv}$ from exploiting a compromised cloud QE outside of the physically secure datacenter perimeter.
Instead, the approach described in Section~\ref{subsec:anti-cuckoo} leverages the cryptographic properties of EPID scheme, without modifying the attestation protocol.

\section{Future Work}
\label{sec:limitations_of_trusdn}
Along with security guarantees, the use of Intel SGX imposes limitations on $\mathsf{TruSDN}$.
Further performance evaluation may be done once software and hardware support for Intel SGX becomes available; 
moreover, we note several security limitations.
Controlled-channel attacks~\cite{xu:2015} are a novel type of side-channel attacks allowing the OS to extract data from protected applications.
They were successfully applied to ``Haven''~\cite{baumann:2014} and $\mathsf{TruSDN}$ could also be vulnerable; 
however, we explicitly excluded such attacks from the adversary model.
Known mitigations are: rewriting applications to decouple memory access patterns from sensitive data, prohibiting paging by the OS, or obfuscating memory access patterns~\cite{xu:2015}.
Another limitation stems from the reliance on the platform vendor, which could leak $QE^{sk}$, to create a ``deniable back-door" and allow \textit{person-in-the-middle} attacks on attestation~\cite{rutkowska:2013.b}.
This challenge remains unaddressed.

In future work we aim to integrate TruSDN with other approaches to cloud infrastructure security, such as in~\cite{paladi:2016}, to provide a complete framework for secure cloud infrastructure deployments in the given adversarial model.
\section{Conclusion}
\label{sec:conclusion}
We described, implemented and evaluated $\mathsf{TruSDN}$ -- a framework for bootstrapping trust in SDN infrastructure.
It isolates network endpoints and switches in SGX enclaves, remotely attests their integrity, and establishes secure communication channels.
We leveraged the principles of SDN to introduce \textit{ephemeral flow-specific PSK} distributed at flow creation, which reduce the overhead of key derivation and reduce the total time to establish protected channels, at the cost of a minor delay in the flow rule installation.
Finally, we leveraged the properties of the EPID scheme to propose an improved approach to prevent cuckoo attacks.

\section{Acknowledgements}
\label{sec:acknowledgements}
This research has been performed within 5G-ENSURE project (www.5GEnsure.eu) and received  funding  from  the European Union's Horizon 2020 research and innovation programme under grant agreements No 671562 and No 644814.

\bibliographystyle{splncs03}

\bibliography{SDN}
\end{document}